\documentclass[a4paper,12pt]{book}
\usepackage[utf8]{inputenc}
\usepackage{graphicx}
\usepackage{multirow}
\usepackage{french}
\usepackage{pdfpages}
\usepackage{natbib}

\usepackage[T1]{fontenc}
\usepackage{titlesec}

\usepackage{setspace}
\newenvironment{changemargin}[2]{%
	\begin{list}{}{%
			\setlength{\topsep}{0pt}%
			\setlength{\leftmargin}{#1}%
			\setlength{\rightmargin}{#2}%
			\setlength{\listparindent}{\parindent}%
			\setlength{\itemindent}{\parindent}%
			\setlength{\parsep}{\parskip}%
		}%
		\item[]}{\end{list}}

\usepackage{colortbl}
\titleformat
{\chapter} 
[display] 
{\bfseries\Large\itshape} 
{} 
{0.5ex} 
{
	\rule{\textwidth}{1pt}
	\thechapter.
	\vspace{1ex}
	\centering
} 
[
\vspace{-0.5ex}%
\rule{\textwidth}{0.3pt}
] 

\titleformat
{\part} 
[display] 
{\bfseries\Large\itshape} 
{} 
{0.5ex} 
{
	\rule{\textwidth}{1pt}
		\rule{\textwidth}{1pt}
		\thepart
			\vspace{1ex}
	\centering
} 
[
\vspace{-0.5ex}%
\rule{\textwidth}{0.3pt}
] 

\begin{document} 
	
		\begin{titlepage}
		\begin{center}
			\vspace*{1cm}
			
				\includegraphics[width=0.4\textwidth]{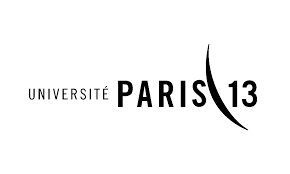}
			\includegraphics[width=0.4\textwidth]{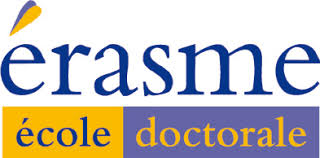}
			
			{THESE DE DOCTORAT}\\
			DE\\
			L'Universit\'e de Paris XIII-Sorbonne Paris Cité\\
			Ecole Doctorale Erasme - Sciences économiques et de gestion\\
			 	CEPN (UMR-CNRS 7234)\\
				\vspace{0.8cm}
			{ M. Sébastien VALEYRE 
			}
			
			\vspace{0.8cm}
			
			
			\vspace{0.8cm}
			{\Large Modélisation fine de la matrice de covariance/corrélation des actions}\\
				\vspace{0.8cm}
			\begin{footnotesize}	
			Thèse présentée et soutenue le 29 Mai 2019, à 10h, à la Maison du Poumon, 66 boulevard Saint-Michel, Paris. Composition du Jury:\\

			\begin{tabular}{ l c r }
				M. Prigent, Jean-Luc & Professeur, Université de Cergy-Pointoise&Président du jury  \\
				M. Malevergne, Yannick & Professeur, Université Paris 1 & Rapporteur  \\
					M. Maillet, Bertrand & Professeur, EM Lyon & Rapporteur  \\
								M. Courtault, Jean-Michel & Professeur, Université Paris 13 & Examinateur  \\
									Mme. Rivals, Isabelle & Maitre de conférence, E.S.P.C.I. & Examinateur  \\
									
							M. Aboura, Sofiane & Professeur, Université Paris 13 & Directeur de thèse  \\
			
			\end{tabular}
			\end{footnotesize}
			
		\end{center}
	\end{titlepage}
	



\frontmatter

\mainmatter
\begin{center}
	\vspace*{1cm}
\large{Résumé}
	\end{center}
\begin{footnotesize}
Une nouvelle méthode a été mise en place pour débruiter la matrice de corrélation  des rendements des actions en se basant sur une analyse par composante principale sous contrainte en exploitant les données financières. Des portefeuilles, nommés ``Fundamental Maximum variance portfolios'', sont construits pour capturer de manière optimale un style de risque défini par un critère financier (``Book'', ``Capitalization'',etc.). Les vecteurs propres sous contraintes de la matrice de corrélation, qui sont des combinaisons linéaires de ces portefeuilles, sont alors étudiés. Grâce à cette méthode, plusieurs faits stylisés de la matrice ont été mis en évidence dont: i) l'augmentation des premières valeurs propres  avec l'échelle de temps de 1 minute  à plusieurs mois semble suivre la même loi pour toutes les valeurs propres significatives avec deux régimes; ii) une loi ``universelle'' semble gouverner la composition de tous les portefeuilles ``Maximum variance''. Ainsi selon cette loi, les poids optimaux seraient directement proportionnels au classement selon le critère financier étudié; iii) la volatilité de la volatilité des portefeuilles ``Maximum Variance'', qui ne sont pas orthogonaux, suffirait à expliquer une grande partie de la diffusion de la matrice de corrélation; iv) l'effet de levier (augmentation de la première valeur propre avec la baisse du marché) n'existe que pour le premier mode et  ne se généralise pas aux autres facteurs de risque. L'effet de levier sur les beta, sensibilité des actions avec le ``market mode’’, rend les poids du premier vecteur propre variables.

\vspace{1.8cm}
\end{footnotesize}
\textbf{Mots clefs}: corrélation, filtre, diagonalisation sous contrainte, modèle multifactoriel, portefeuilles optimaux, gestion d'actifs, diffusion

\newpage 
A mon épouse, Léa,
A ma fille Gabrielle,
A mes parents, Françoise et Dominique

\begin{center}
	\vspace*{1cm}
	\large{Remerciements}
\end{center}

Je souhaite ici rendre hommage et exprimer ma profonde gratitude à tous ceux qui, de près ou de loin, ont contribué à la réalisation de cette thèse.
Mes remerciements s’adressent tout d’abord à mon Directeur de thèse, le Professeur Sofiane Aboura. Tout au long de ce travail, il a su m’apporter  une disponibilité, une écoute, une confiance, des conseils précieux et avisés à la hauteur de ses compétences et de ses  qualités humaines.
J’adresse également tous mes remerciements aux Professeurs Bertrand Maillet et Yannick Malevergne, de l’honneur qu’ils m’ont fait en acceptant d’être rapporteurs de cette thèse ainsi qu'aux Professeur Jean-Michel Courtault, Jean-Luc Prigent et Isabelle Rivals de l’honneur qu’ils m’ont fait en acceptant d'être examinateur. 
Je remercie aussi mes collègues, l'équipe de recherche de John Locke, Denis Grebenkov, Stanislav Kuperstein, Maxime Beucher, Edouard Limouse et Frédéric Herbette. Sans eux, ce travail n’aurait probablement pas vu le jour. Leur aide et soutien ont été déterminants. J'ai aussi une pensée pour Thomas Dionysopoulos, avec lequel, suite à nos discussions en janvier 2015,  j’ai commencé à réfléchir à l’opportunité de  publier des articles sur la matrice de corrélation. 

Enfin, sur un plan plus personnel, l’amour que je porte à mon épouse, Léa, et à ma fille, Gabrielle, m’a permis de surmonter les étapes difficiles.
J’ai aussi une pensée pour les personnes qui ont été déterminantes dans ma reconversion professionnelle il y a 12 ans de l’énergie nucléaire à la finance de marché dont cette thèse est l’un des jalons. Je me souviens notamment de ma discussion avec le Professeur Jean-Philippe Bouchaud, en mai 2006, au jardin du Luxembourg, qui a été un moment crucial. Il a été très bienveillant en me donnant des bons conseils: il ne fallait surtout pas reprendre mes études dans les produits dérivés, les calculs stochastiques et les équations différentielles, il fallait au contraire reprendre mes études axées sur la recherche, l'empirisme et les inefficiences de marché, sujet qui n'était pas encore très à la mode avant la crise financière. Je me souviens, par sa clairvoyance, qu'il avait déjà anticipé la crise des subprimes de 2007-2008 en me disant que selon lui les hypothèses liées à la valorisation des produits financiers dérivés complexes étaient devenues surréalistes. J'ai toujours été impressionné par sa simplicité, sa disponibilité et sa gentillesse vu sa réussite exceptionnelle. Je me souviens à cette époque du nombre de journées que j’ai passées à lire son livre et ses nombreux articles de recherche qui m’ont permis de former mes premières intuitions par rapport au fonctionnement des marchés financiers. J’ai aussi une pensée émue pour les Professeurs Jacques Prost et Pierre-Gilles de Gennes, tout aussi simples, bienveillants et exceptionnels, qui ont permis probablement en me recommandant, que l’université dauphine m’accepte en tant qu’étudiant, à 28 ans, avec mon parcours atypique au master de finance. J’ai aussi une pensée pour Joel Benarroch et François Bonnin qui m’ont appris le métier de gérant de fonds discrétionnaire et systématique. J’ai aussi une pensée pour les seuls professeurs de l’ESPCI qui ont vraiment cru en moi, Isabelle Rivals et Léon Personnaz, qui m'ont formé aux statistiques pendant un an dans leur laboratoire. Ils ont eu beaucoup d’influence sur mes travaux. J’ai aussi une pensée pour mes amis les plus proches Pierre, Alexandre, Charles-Henri, Olivier, Emmanuel et Antoine et pour mes parents et mes frères et sœurs, qui ont cru en moi, et m'ont toujours soutenu dans cette reconversion professionnelle et je les remercie aussi.

\if false

Une nouvelle méthode a été mise en place pour débruiter la matrice de corrélation  des rendements des actions en se basant sur une analyse par composante principale sous contrainte en exploitant les données financières. Des portefeuilles, que j'ai nommés ``Fundamental Maximum variance portfolios'', sont construits pour capturer de manière optimale un style de risque défini par un critère financier (``Book'', ``Capitalization'',etc.). Les vecteurs propres sous contraintes de la matrice de corrélation, qui sont alors des combinaisons linéaires de ces portefeuilles, sont alors étudiés. Grâce à cette méthode, plusieurs faits stylisés de la matrice ont été mis en évidence dont: \begin{itemize}
	\item l'augmentation des premières valeurs propres  avec l'échelle de temps de 1 minute  à plusieurs mois semble suivre la même loi pour toutes les valeurs propres significatives avec deux régimes : de quelques secondes à quelques minutes, l’augmentation des corrélations vient d’un effet retard alors que de quelques jours à plusieurs mois l’augmentation des corrélations vient  d’un manque de liquidité du marché et du comportement moutonnier des agents.
	\item une loi ``universelle'' semble gouverner la composition de tous les portefeuilles ``Maximum variance''. Les poids optimaux seraient directement proportionnels au classement selon le critère financier étudié. Le portefeuille ainsi construit capturerait mieux le risque recherché tout en minimisant le risque spécifique. 
	\item La volatilité de la volatilité des portefeuilles ``Maximum Variance'', qui ne sont pas orthogonaux, suffirait à expliquer une grande partie de la diffusion de la matrice de corrélation. Une des propriétés reproduite est la forme de la distribution des valeurs propres des variations de la matrice de corrélation, qui ne suit pas la loi demi cercle de Wigner et qui  n'est pas capturée par des modèles standards de la littérature qui sont tous dérivés du processus de Wishart. La diffusion de la distribution des valeurs propres des variations de la matrice de corrélation s'expliquerait plus par la diffusion des vecteurs propres de la matrice de corrélation que par la diffusion des valeurs propres de la matrice de corrélation.  Les modèles stochastiques standards de la matrice de covariance semblent minimiser fortement les plus grandes valeurs propres des variations de la matrice de corrélation et donc semble sous-estimer les chocs de volatilité et la probabilité de pertes extrêmes de certaines stratégies d’investissement.
	\item L'effet de levier (augmentation de la première valeur propre avec la baisse du marché) n'existe que pour le marché et  ne se généralise pas aux autres facteurs de risque. L'impact de l'effet de levier sur les beta, sensibilité des actions avec le ``market mode’’,  a aussi été modélisé ainsi que son élasticité. Quand une action surperforme, son beta va baisser. Quand le marché baisse, les corrélations vont augmenter. Quand la volatilité de l'action augmente plus que celle du marché,  le beta de l'action augmente aussi. Je montre qu'il est important de tenir compte de ces effets pour mesurer les beta sans biais et construire des portefeuilles ``market neutral'' c'est à dire complètement insensibles aux variations du marché. 
\end{itemize}
Ces différents faits stylisés doivent être pris en compte pour modéliser proprement la matrice de corrélation ce qui est primordiale dans la gestion de portefeuille et la gestion des risques. Cela est aussi utile pour l'Asset Pricing et identifier que non seulement les entreprises qui rémunèrent bien leurs employés ont tendance à partager un risque en commun mais ont aussi tendance à surperformer.
\fi
\newpage 
\begin{center}
	\vspace*{1cm}
	\large{Avant-Propos}
\end{center}

Cette thèse a débuté en janvier 2016 dans le laboratoire CEPN (UMR 7234, CNRS) de l'université  Paris-XIII. Le travail de recherche a été réalisé chez John Locke Investments, société de gestion indépendante et à taille humaine (15 salariés), pour laquelle j'ai continué à travailler à pleins temps en tant que chercheur et gérant des fonds systématiques John Locke Equity Market Neutral et John Locke Smart Equity. J'ai ainsi pu profiter de mon expérience concrète des marchés financiers pour adapter mes modèles à la réalité. Aussi j'ai dû me concentrer sur des modèles qui devaient avoir un intérêt certain pour la gestion d'actifs et les deux fonds que je gère. Modéliser la matrice de corrélation des actions est clef chez John Locke Investments. Ainsi les portefeuilles optimaux pour faire du trend following se basent uniquement sur l'exploitation de la matrice de corrélation qu'il faut maitriser, nettoyer, inverser, modéliser très proprement pour pouvoir amplifier les faibles autocorrélations en performances robustes. Ainsi les compétences de certains gérants peuvent très bien se limiter à la bonne modélisation de la matrice de corrélation. Les papiers de recherche devaient aussi être pratiques et constituer un support intellectuel pour convaincre les clients des fonds du fondement scientifique de mes modèles de gestion.

 Trois papiers ont été présentés lors des conférences en 2016 (Liège, Belgique), en 2017 (Valence) et en 2018 (Paris) de l'AFFI et un quatrième sera présenté lors de la conférence en Juin 2019 (Laval, Quebec): 
 \begin{itemize}

	\item  le papier ``Emergence of Correlation between Securities at Short Time Scales''  a été présenté au 35th International Conference of the French Finance Association à l’ESCP à Paris du 20 au 24 mai 2018. 
	Le papier est présenté en premier au chapitre \ref{emergence} car il explique l'origine physique des corrélations entre actions;
	\item  le papier ``Fundamental Market Neutral Maximum Variance Portfolios''   a été soumis en janvier 2019 au 36th International Conference of the French Finance Association à Laval au Quebec. Le papier justifie la méthodologie utilisée dans la thèse pour débruiter la matrice de correlation. Le papier est à cheval entre plusieurs spécialités (model factoriels, matrice aléatoires, Asset Pricing) et doit être restructuré et découpé en plusieurs projets pour être publiable. Le papier est présenté au chapitre \ref{maxvar}.
	\item le papier ``The Reactive Beta Model'' a été présenté au 34th International Conference of the French Finance Association à Valence le 31 mai et 1er et 2 juin 2017. 
	Le papier est présenté au chapitre \ref{beta};
	\item  le papier ``Should Employers Pay Better their Employees? An Asset Pricing Approach''  a été présenté au	33rd International Conference of the French Finance Association à HEC-Management School of the University of Liege du 23 au 25 mai 2016. Le papier est présenté dans le manuscrit au chapitre \ref{remuneration} comme une application.

\end{itemize}

\tableofcontents
\part{Introduction Générale}
\chapter{Introduction}

La matrice de corrélation des rendements des actions est nécessaire à l'analyse du risque d'un portefeuille. Une modélisation fine est nécessaire pour construire les portefeuilles optimaux robustes (maximiser le gain potentiel tout en minimisant le risque). Les mesures empiriques de la matrice de corrélation sont bruitées du fait d'un nombre trop faible de rendements indépendants et homoscedastiques disponibles et d'un nombre trop grand d'actions. Ainsi il est courant de devoir mesurer les corrélations entre 500 actions ou plus avec beaucoup moins d'un an d’historique\footnote{ce qui correspond à moins de 250 rendements journaliers qui ne sont que très approximativement gaussiens} afin de pouvoir supposer que les corrélations restent à peu près constantes sur cette période. L’échantillon se réduit encore lorsqu’on s’intéresse aux corrélations des rendements mensuels voire annuels, qui importent le plus pour les investisseurs. Les autocorrélations des rendements sont faibles mais suffisantes pour déformer la matrice selon l’horizon de temps et transformer des facteurs de risque négligeables à l'horizon de la journée en significatif à l'horizon du mois pour un gérant.  Les mesures empiriques, qui se basent sur un échantillon trop petit, capturent des corrélations fallacieuses. Ces mesures fallacieuses peuvent résulter en portefeuilles qui semblent sans risque dans l'échantillon utilisé pour mesurer la matrice mais risqués dans un autre échantillon. La moindre optimisation de portefeuille, qui cherche à minimiser le risque pour une même rentabilité espérée, va forcément privilégier les portefeuilles qui semblent sans risque ou très peu risqués ``in the sample'' et il en résulte un manque de robustesse. Les vraies corrélations sont réputées être de plus très variables en fonction du temps. Ainsi quand le marché est stressé, les investisseurs se mettent à paniquer et les corrélations ont tendance à augmenter, si bien que toutes les actions sont entrainées par les mouvements des indices. Lorsqu'un facteur de risque devient majeur, lorsqu'un évènement inattendu  survient, alors toutes les actions qui capturent ce facteur de risque vont se corréler brusquement. Lorsqu'une action sous-performe ou surperforme, ses corrélations avec les autres actions vont changer. Ainsi on peut parler de corrélations non linéaires car les corrélations dépendent des trajectoires de chaque action mais une grande partie des variations semblent complètement stochastiques.  Une grande difficulté est donc de mesurer les corrélations de ``population'' sans erreur et sans retard. Une autre difficulté est aussi de prévoir comment la matrice risque de varier.   Malgré les enjeux, de nombreux faits stylis\'es de la matrice de corrélation, qui sont noyés dans le bruit, restent pourtant encore à découvrir. Cette thèse, qui cherche à mettre en évidence plusieurs faits stylisés, remplit un vide dans la littérature académique et fait le lien entre plusieurs disciplines entre mathématiques (processus stochastique), finance (modèles multifactoriels), gestion d'actifs (portefeuilles optimaux), econophysique (matrice aléatoire) et économie (Asset Pricing).

Je me suis d'abord intéressé à l'origine des corrélations des rendements des actions: j'ai ainsi modélisé l'émergence des corrélations des rendements  des actions européennes et américaines de 2000 à 2017 sur des \'echelles courtes (de 1 minute à 1 jour) grâce à un modèle de retard inspir\'e de la microstructure. A des échelles très courtes, de l'ordre de la seconde, les corrélations sont nulles puis elles apparaissent et augmentent avec l'échelle de temps. L'émergence des corrélations est en fait la conséquence de l'impact des transactions, qui se matérialisent entre les actions similaires via des algorithmes de trading. J'ai mis en place, dans le chapitre \ref{emergence}, un mod\'ele de retard qui reproduit très bien l'effet d'\'echelle mesur\'e qui permet d'extrapoler la vision de la matrice des rendements  1 minute à la journ\'ee.

Pour identifier la structure et la dynamique des valeurs propres et vecteurs propres, j'ai mis en place, dans le chapitre \ref{maxvar}, une m\'ethodologie bas\'ee sur l'analyse par composante principale contrainte qui permet de débruiter la matrice en tirant bénéfice des informations financières, comme, par exemple, le ratio entre la valeur comptable et la valeur de marché de l'action (``Book''), la valeur capitalistique de l'action (``Capitalization'') ou de nombreux autres ratios financiers.  L'analyse par composante principale appliquée aux rendements de l'ensemble des actions revient à diagonaliser la matrice de corrélation des rendements. La matrice de corrélation est préférable à la matrice de covariance pour éviter un biais vers les actions les plus volatiles. La diagonalisation  permet d'identifier les portefeuilles d'actions decorrélés les uns des autres qui génèrent le plus de volatilité pour un même investissement (mesurée exactement par la volatilité du portefeuille obtenue sans tenir compte des corrélations entre les actions). Les rendements de ces portefeuilles particuliers permettent de modéliser simplement les mouvements principaux du marché: ces portefeuilles particuliers sont réputés proches des combinaisons très bruitées de stratégies de base  (les indices pondérés par les capitalisations, les indices sectoriels, les indices investis sur les petites capitalisations, les indices investis sur les entreprises de croissance, les indices ``Min Variance'' investis sur les actions peu volatiles, les indices investis sur les entreprises ``Value'', etc.).  Les variances de ces portefeuilles lorsqu'ils sont normalisés  sont proportionnelles aux valeurs propres. Les corrélations entre actions sont quasiment toutes positives et rend l'identification du premier vecteur propre plutôt aisée: le premier vecteur propre est très significatif. Il reste proche du portefeuille investi sur chacune des actions avec une valeur propre de l'ordre de 100\footnote{proche de la corrélation moyenne de l'ordre de 0.4 au carré multipliée par le nombre d'action de l'ordre de 500 dans mon cas}.  Les autres vecteurs propres ont des valeurs propres beaucoup plus petites (inférieure à 20) et représentent des portefeuilles ``long/short'' et ``market neutre'' d'abord plutôt sectoriels puis plutôt de style. Cependant l'instabilité de la matrice de corrélation associée au bruit de mesure rend difficile l'interprétation des vecteurs propres ``long/short'' mesurées. Ainsi d'un côté, on devrait réduire la profondeur sur laquelle on mesure la corrélation pour espérer une certaine stabilité des corrélations sur la période de mesure,  mais, de l'autre coté, on devrait augmenter la période et la fréquence pour réduire le bruit de mesure. 

Pour filtrer le bruit de mesure, inhérent à l'analyse par composante principale, j'ai contraint l'analyse au sous espace des facteurs de risque principaux, déjà identifi\'es dans la litt\'erature, dont j'ai optimis\'e la construction. J'ai inclus les facteurs de styles principaux (``Momentum'', ``Capitalization'', ``Quality'', etc.) et les facteurs de risque sectoriel. Les facteurs principaux optimisés ont été nommés ``Fundamental Maximum variance market neutral portfolios'', car la variance de leurs rendements a été optimis\'ee par construction. Ces facteurs peuvent aussi être directement utiles dans l'industrie de la gestion d'actifs, car ils optimisent théoriquement le gain ajusté du risque des primes de risque alternatives, qui sont devenues des véhicules d'investissement très populaires. Aussi grâce à l'optimisation, j'ai pu relier les valeurs propres sous contraintes débruit\'ees aux  valeurs propres bruit\'ees de la matrice. Cela m'a permis de débruiter la matrice de corrélation et de caract\'eriser finement une loi universelle, selon laquelle, les poids optimaux des facteurs de risque seraient uniformément distribués pour tous les critères financiers ce qui est particulièrement intriguant (on aurait attendu une distribution gaussienne et non uniforme plus logique pour obtenir des vecteurs propres aléatoires). Cette loi universelle a des conséquences importantes dans l'Asset Pricing: la norme dans cette discipline est de construire des portefeuilles ``long/short'' investis à l'achat sur le premier quintile selon le critère financier étudié et à la vente sur le dernier quintile. Si le portefeuille capture une performance significativement différente de zéro alors une anomalie de marché est identifiée. Une construction plus optimale du portefeuille avec une règle d'investissement linéaire au lieu de la marche en escalier peut aider à obtenir des performances plus significatives pour les petites anomalies.

 Le filtre du bruit de mesure m'a aussi permis de caractériser finement la dynamique de la matrice de corrélation et d'identifier notamment les violents changements des valeurs propres (la valeur propre du premier mode peut passer de 200 à 30 soit une variation de corrélation moyenne de 0.5 à 0.05 en quelques mois seulement). Ce violent changement peut être mod\'elis\'e par l'effet de levier (corrélation négative entre les rendements et les volatilités) pour la premi\'ere valeur propre. J'ai aussi v\'erifi\'e que les premiers vecteurs propres s'investissaient sur les facteurs fondamentaux de risque les plus risqués qui sont différents selon les périodes. Selon les crises, il peut s'agir du secteur IT, du secteur de la finance, du secteur de l'énergie, des REITs ou des entreprises exposées à la dette. Les entreprises qui sont peu sensibles aux variations de l'indice, et celles qui constituent les composants du facteur ``Momentum’’, restent très représentées dans le deuxième et troisième vecteurs propres. Les facteurs ``Capitalization'' et ``Book'' de Fama et French sont très peu représentés dans les premiers vecteurs propres de la matrice de corrélation.

La première application de la méthodologie que j'ai introduite et qui permet de débruiter la matrice de corrélation a consisté à étendre l'étude de l'effet d'\'echelle sur les valeurs propres de 1 minute à 1 journée sur des échelles de temps plus longues entre 1 jour et plusieurs mois.  Les corrélations continuent d'augmenter avec l'échelle de temps. Cela explique par exemple que la norme  dans l'Asset Pricing est de se baser sur les rendements mensuels pour estimer les corrélations. En effet, même si l'utilisation des rendements journaliers donnerait des résultats plus robustes, les chercheurs dans l'Asset Pricing préfèrent travailler avec les rendements mensuels, car les corrélations sont réputées plus fortes lorsqu'elles sont mesurées à partir de rendements mensuels qu'à partir des rendements journaliers à cause d'un effet d'echelle que les chercheurs redoutent. La réduction de la matrice de corrélation dans le sous espace généré par les portefeuilles fondamentaux ``Maximum variance'' permet de confirmer cette crainte avec des mesures significatives. Les corrélations ont tendance à continuer à augmenter sur des horizons de temps plus long. Ce phénomène est expliqué grâce à un modèle d'autocorrélation, qui permet de reproduire l'effet de manque de liquidit\'e du march\'e. L'illiquidité crée de l'inertie et fait qu'un mouvement de marché dure et peut être prolongé par le comportement moutonnier des investisseurs. Les autocorrélations, introduites dans le chapitre \ref{emergence2}, apparaissent plus robustes que les anomalies non conditionnelles pas toujours significatives telles qu'identifiées dans l'Asset Pricing. Ces anomalies se matérialisent par des primes de risque alternatives pour justifier les incohérences avec le modèle d'évaluation des actifs financiers (MEDAF ou CAPM en anglais), selon lequel, les primes de risque ne doivent dépendre que du beta, sensibilité de l'action avec les variations de l'indice.

La deuxième application de de la méthodologie que j'ai introduite et qui permet de débruiter la matrice de corrélation a consisté à caractériser la dynamique de la matrice de corrélation qui est importante à modéliser pour estimer les risques. En effet la matrice de corrélation de population peut changer et cela peut représenter un risque. Le problème est que la matrice est déjà tellement bruitée qu'espérer mesurer ces changements est illusoire, si bien que les modèles stochastiques théoriques ne peuvent pas facilement être validés empiriquement. Dans le chapitre \ref{diffusion}, j'ai réussi à faire plusieurs mesures grâce à la méthodologie qui permet d'utiliser les informations financières pour réduire la taille de la matrice de corrélation et grâce à l'utilisation des rendements 5 minutes. J'ai ainsi mis en évidence certains faits stylisés de la diffusion de la matrice de corrélations très mal connus et très mal reproduits par les modèles standard issus de Wishart, comme la distribution des valeurs propres des incréments de la matrice de corrélations des actions (il s'agit ici précisément de la matrice de corrélation des actions sous sa forme réduite dans le sous espace des 24 facteurs fondamentaux Maximum Variance pour éliminer le bruit de mesure). L'étude de cette distribution permet de caractériser  la diffusion des vecteurs propres de la matrice de corrélation des actions. Cette distribution ne suit pas une loi demi-cercle de Wigner mais une distribution avec des queues, qui peuvent être interprétées par la présence de valeurs propres extrêmes. Ces valeurs propres extrêmes expliquent que des corrélations entre actions peuvent changer beaucoup plus brutalement que les modèles classiques ne peuvent le prévoir.  J'ai ainsi modélisé l'instabilité de la matrice de corrélation avec un processus empirique plus réaliste. La diffusion dans la composition des premiers vecteurs propres explique en grande partie la distribution des valeurs propres des incréments de la matrice de corrélation. Cette diffusion s'explique quasiment entièrement par la volatilité de la volatilité des portefeuilles fondamentaux ``Maximum variance''. Les portefeuilles n'étant pas orthogonaux, la volatilité de la volatilité permet de répliquer la diffusion des vecteurs propres tout en supposant les corrélations entre portefeuilles fondamentaux fixes.

Une composante particulière de la diffusion de la matrice a aussi fait l'objet d'une grande attention: la dynamique des poids du premier vecteur propre qui sont liés aux beta, qui est la sensibilité des rendements d'une action avec les indices boursiers, a été analysée en profondeur. Les beta constituent par ailleurs une mesure du risque qui est capitale car ils forment une indication d'un risque systématique qui ne peut pas se diversifier ou s'éliminer pour un investisseur classique. Cela justifie intuitivement que les actions à fort beta doivent rémunérer plus les actionnaires et que les primes de risque doivent être proportionnelles au beta qui est à la base du MEDAF. Aussi les fonds alternatifs, qu’on appelle aussi ``hedge funds’’, ont la capacité à prendre des positions vendeuses avec des ventes à découvert pour neutraliser l'exposition de leur investissement aux variations des indices boursiers. Cela permet de mieux contrôler le risque et de proposer des investissements diversifiant aux épargnants. Pour construire des portefeuilles immunisés contre les variations de la bourse, qu'on appelle ``beta neutre'', il est extrêmement important de se baser sur des mesures fiables et sans biais des betas d'autant plus que certaines stratégies très populaires ont tendances à amplifier les biais de mesure du beta. Cela m'a motivé à mod\'eliser finement l'effet de levier et l'élasticit\'e des beta, qui décrivent aussi la composante du premier vecteur propre de la matrice. Par exemple, lorsqu'une action sous performe, son beta va augmenter. Lorsque la volatilité de l'action augmente plus que les autres, son beta va augmenter aussi. J'ai mis au point, dans le chapitre \ref{beta}, une m\'ethode r\'eactive de la mesure des beta n\'ecessaire pour construire des facteurs fondamentaux beta et secteur neutre moins biaisés et potentiellement mieux valider le MEDAF. Des tests montrent l'intérêt d'un tel modèle par rapport à des méthodes standards (OLS, régression par quantile, DCC GARCH).

Enfin une application concrète de mes travaux, dont la portée peut ne pas se limiter à la gestion d'actifs, met en avant l'intérêt de la m\'ethode que j'ai introduite  en se révélant assez fine pour distinguer  le facteur ``Rémuneration’’ du bruit. Dans le chapitre \ref{remuneration}, je montre que facteur ``Rémuneration’’ s'avère être un facteur de risque commun significatif. Les entreprises qui rémunèrent mieux leurs employ\'es ont un risque en commun. Ces entreprises ont aussi tendance à avoir des meilleures performances. J'ai ainsi découvert une nouvelle anomalie par rapport au MEDAF et aux facteurs de Fama et French qui pourrait avoir une portée managériale voire politique.

Cette thèse peut donc avoir de multiples applications: une meilleure analyse du risque, une optimisation plus robuste d'un portefeuille, une meilleure modélisation des autocorrélations qui sont exploitées par les programmes de trading d'arbitrage de style, une meilleure mesure des anomalies dans l'Asset Pricing, une modélisation plus réaliste de la dynamique de la matrice de corrélations pour évaluer des produits dérivés. Elle peut aussi avoir des implications très concrètes en économie et en management car, par exemple, elle permet de montrer que les entreprises qui rémunèrent bien leurs employés partagent un risque significatif en commun et ont aussi tendance à mieux performer.

\chapter{Revue de la littérature}

Ce travail de recherche s'est articulé autour de six champs disciplinaires  relativement cloisonnés entre plusieurs disciplines finance, économie, éconophysique et mathématiques appliquées.

\section{Gestion de portefeuille}
\label{ptfmng}
La bonne estimation des corrélations des rendements des actions est nécessaire pour l'analyse de risque d'un portefeuille  et pour son optimisation. La bonne compréhension des variations temporelles des corrélations en cours ou potentielles est aussi critique pour la gestion d'un portefeuille et notamment d'un fonds ``market neutre'' qui utilise un fort effet de levier financier et dont l'arbitrage de style est un des moteurs  de performance. L'arbitrage de style est une stratégie de trading qui consiste à investir sur les styles de gestion porteurs. Par exemple si le style de gestion qui consiste à acheter des petites capitalisations et à vendre des grosses capitalisations est profitable, la stratégie va  acheter les petites capitalisations et vendre les grandes. Dans le cas inverse, la stratégie va acheter les grosses et vendre les petites. Plus de 240 styles de gestion ou facteurs de risques profitables ont été publiés dans la littérature scientifique décrites dans la section \ref{Assetpricing}. L'intérêt du market timing ne fait pas consensus (\cite{Lee17, Bender18,Bass17}) et certains préfèrent bénéficier simplement de la diversification. \cite{Miguel17} montrent qu'en pratique il suffit, pour construire un portefeuille, de sélectionner 15 critères financiers significatifs sur plus de 100. Les stratégies de market timing peuvent être complexes. Elles s'appuyent sur des modèles de prévision. \cite{Hodges17} cherchent des prédicteurs des facteurs dans diffèrent régimes économiques et différentes conditions de marché. Ils trouvent que l'utilisation d'une combinaison d'indicateurs sur le cycle économique, la valorisation, la tendance et la dispersion serait plus efficace que l'utilisation d'indicateur individuel. Ainsi \cite{Dichtl18} fabriquent un portefeuille ``long/short'' grâce à la méthode d'optimisation des paramètres introduite par \cite{Brandt09}  en utilisant plusieurs indicateurs de valorisation et de tendance et ils montrent que le market timing permet de surperformer le portefeuille investi équitablement sur les différents styles de gestion dont les primes de risque sont positives. La fragilité de ces résultats vient du risque de surapprentissage. De plus ces stratégies en général peuvent souffrir de chocs de corrélations entre les différents styles de gestion qui peuvent survenir et générer des pics de volatilités. Ainsi les stratégies quantitatives d'habitude non corrélées peuvent se corréler fortement de manière brutale. Cela s'est passé  du 8 au 9 août 2007, quand la plupart des fonds d'arbitrage de style ont subi des pertes très significatives brutalement en même temps (\cite{Stein09}). Lors de cet évènement, nommé plus tard ``quant crash'', la plupart des fonds touchés employaient des stratégies ``market neutre'' quantitatives sans exposition au marché ce qui remet en question leur statut ``market neutre'' (\cite{Khandani11}). Il semble en fait que trop de gérants étaient investis sur les même ``crowded'' stratégies avec trop de levier et qu'ils ont tous voulu réduire leurs positions en même temps au même signal. De tels risques de ``crowding'' affectent une grande variété de stratégies, comme le style ``Momentum'' (acheter les actions qui ont surperformé et vendre les actions qui ont sous performé) car ils ne dépendent pas d'estimation indépendante des valeurs fondamentales des entreprises (\cite{Hong16,Stein09}).  

Des centaines de milliards de dollars sont aussi gérées directement  en utilisant  l'optimisation Mean-Variance introduite par \cite{Markowitz52}  en préférant se baser sur des hypothèses simples concernant les espérances des rendements. La valeur ajoutée des gérants viendrait seulement d'une modélisation plus adaptée de la matrice de corrélation et d'une bonne capacité à exécuter les ordres en minimisant l'impact de marché. Ainsi le portefeuille ``Min Variance'' suppose que les espérances des rendements sont toutes identiques et que la matrice de corrélation peut se modéliser simplement, par exemple, avec un modèle à un facteur (\cite{Clarke13}). Le portefeuille ``Max Diversification'' introduit par \cite{Choueifaty08} suppose que les espérances sont proportionnelles au risque. Ces deux derniers portefeuilles nécessitent d'inverser la matrice de corrélation ce qui peut poser problème si la matrice n'est pas proprement modélisée. Le portefeuille ``Equal-Risk Contribution'' introduit par \cite{Maillard10} est moins sensible aux bruits de mesure et est donc plus robuste mais n'est plus théoriquement optimal. \cite{Benichou16} introduisent le portefeuille ``Agnostic Risk Parity''. Les espérances ne sont plus forcément positives mais dépendent des rendements passés. Le portefeuille dépend alors de l'inverse de la racine carré de la matrice de corrélation multipliée par des signaux qui représentent des indicateurs techniques des tendances. Ce portefeuille ``trend following'' alloue le même risque sur chaque vecteur propre de la matrice de corrélation.

\section{Econophysique}
Aujourd'hui les performances des différents styles de gestion et les performances sectorielles sont très suivies par tous les acteurs du marché qui ne se contentent plus d'avoir une vue binaire (le marché va-t-il monter ou baisser?) et s'auto-alimentent par un phénomène d'effet moutonnier très bien décrit dans la littérature (\cite{Guedj05,Michard05,Cont00,Wyart07,Lux99}). Ainsi quand tel ou tel style de gestion chute, les acteurs vont le vendre en même temps et accentuer sa chute. La moindre nouvelle macroéconomique va impacter les indices mais aussi les autres facteurs de risque. Quand la Réserve fédérale des États-Unis se dit prête à augmenter les taux d'intérêt, le facteur levier (vente d'actions endettées, achat d'actions peu endettées) sera joué puis d'autres facteurs seront entrainés. \cite{Benzaquem16}  mettent en évidence le lien entre le trading et la matrice de corrélation en partant de la microstructure et du cross impact des transactions sur les prix.  Les corrélations ne décriraient que l'interaction entre actions par le jeu des traders. Les corrélations sont aussi réputées pour augmenter avec l'échelle de temps: les rendements mensuels sont plus corrélés que les rendements journaliers qui sont plus corrélés que les rendements 1 minutes (\cite{Epps79}).  \cite{Bouchaud09} avaient déjà proposé dans la partie ``some open problem'' une piste (les rendements de l'action i n'impactent pas instantanément les rendements de l'action j mais avec un certain retard) pour expliquer la dépendance des corrélations à la fréquence  mais ne l'avait pas développé. 

\section{Modèles multifactoriels}
\label{mutifact}
Depuis l'article majeur de \cite{Markowitz52}, l'optimisation  « Mean Variance » est devenue une méthode rigoureuse pour construire un portefeuille d'investissement. Deux ingrédients fondamentaux sont nécessaires: les espérances des rendements de chaque action et la matrice de covariance des rendements. L'estimation de la matrice de covariance a toujours été un sujet important. La méthode de base se contente d'agréger les rendements historiques et de calculer leurs covariances historiques. Malheureusement cela crée des problèmes bien documentés (\cite{Jobson80}). Pour l'expliquer simplement, quand le nombre d'actions est grand devant le nombre d'observations disponibles, ce qui est généralement le cas, la matrice de corrélation historique comporte beaucoup d'erreurs. Cela implique que les coefficients les plus extrêmes prennent des valeurs extrêmes non pas à cause de la réalité mais à cause d'erreurs extrêmes. Invariablement les optimisations de portefeuille vont miser leurs plus gros paris sur ces erreurs extrêmes ce qui rendra l'optimisation extrêmement non fiable. \cite{Michaud89} appelle ce phénomène ``error-optimization''. De manière alternative on peut considérer une estimation avec beaucoup de contraintes, comme le « single-factor model » de \cite{Sharpe63}. Ces estimateurs de la matrice de corrélation contiennent d'un côté  peu d'erreurs mais de l'autre beaucoup d’erreurs de spécification et de biais. Une alternative est le ``Shrinkage'' qui consiste à un mélange entre l'estimation sans contrainte et l'estimation avec la contrainte (\cite{Ledoit03,Ledoit12}). L'APT (``Arbitrage Pricing Theory'') de \cite{Ross76} a généré un intérêt croissant dans les modèles multifactoriels. Ainsi le standard de l'industrie de la gestion d'actifs est d'utiliser des modèles multifactoriels.  Quelques entreprises, comme APT, Barra et Axioma (\cite{Barra98})  qui sont devenues incontournables dans l'industrie de la gestion d'actifs,  proposent à leurs clients des matrices de covariances qui s'adaptent mieux aux optimisations de portefeuille. Ces sociétés ont été accusées d'être à l'origine du ``quant crash'' de 2007, déjà mentionné dans la section \ref{ptfmng}, car elles favorisaient le ``crowding'' en fournissant les même facteurs de risque à tous les gérants. Ces méthodes se basent sur des modèles multifactoriels fondamentaux combinant  une cinquantaine de facteurs sectoriels et d'autres risques. Ces facteurs utilisent le rendement des portefeuilles associés à certains critères financiers observables tel que le ``Dividend Yield'', le ``Book to Market'' ratio ou les secteurs d'appartenance. Une autre approche est d'utiliser les facteurs statistiques issus de l'analyse par composante principale, qui est décrite dans la section \ref{acp}, avec un nombre total de facteurs de l'ordre de 5. \cite{Connor95} montre que les modèles multifatoriels ``fondamentaux'' permettent d'expliquer 42\% ($R^2=42\%$ étant le pouvoir explicatif du modèle) des rendements alors qu'une simple analyse par composant principale  sur 5 facteurs  explique deja 39\%. \cite{Connor95} trie les facteurs selon leur pouvoir explicatif. Les secteurs permettent d'augmenter de 18\%, puis le facteur ``Low Volatility'' (proche du facteur ``Low Beta'')   augmente le $R^2$ de 0.9\% puis les facteurs ``Momentum'', ``Capitalization'', ``Liquidity'', ``Growth'', ``Earning'', augmentent de moins de 0.8\%. Puis il reste par ordre d'importance décroissant des facteurs plutôt mineurs: le ``Book to market'',  le ``Earning Variability'', le ``Leverage'',  l'investissement à l'étranger, le coût du travail et enfin le ``Dividend Yield''. Toutefois la sélection des facteurs nécessaires et le choix du nombre a fait l'objet de nombreuses controverses (\cite{Roll80,Roll84,Dhrymes84,Luedecke84,Trzcinka86,Conway88,Brown89}). \cite{Connor93} proposent une méthodologie simple pour estimer le nombre de facteurs significatifs: si le rajout d'un facteur ne réduit pas significativement le carré du résidu alors le facteur n'est pas considéré comme significatif. La plupart des études académiques se base sur une analyse historique depuis 1967 en exploitant la base de données du centre de recherche des prix des actions (Center of Research in Security Prices). Cette base de données regroupe principalement les actions cotées à la bourse du New-York Stock Exchange depuis 1926.

\section{Analyse par composante principale}
\label{acp}
L'analyse par composante principale (ACP) prend sa source dans un article de Karl Pearson publié en 1901. Encore connue sous le nom de transformée de Karhunen-Loeve ou de transformée de Hotelling, l'ACP a été de nouveau développée et formalisée dans les années 1930 par Harold Hotelling. La puissance mathématique de l'économiste et statisticien américain le conduira aussi à développer l'analyse canonique, généralisation des analyses factorielles dont fait partie l'ACP. Les champs d'application sont aujourd'hui multiples, allant de la biologie à la recherche économique et sociale, et plus récemment le traitement d'images.

La théorie de la matrice aléatoire, dont la distribution des valeurs propres obtenues par l'ACP suit la loi de Mar\v{c}enko-Pastur pour les grandes matrices, modélise les bruits de mesure des corrélations et montre que les petites valeurs propres en dessous d'une valeur propre critique sont sous estimées et ne sont pas significatives  (\cite{Laloux98, Plerou99,  Plerou02, Potters05, Wang11}). \cite{Bun16} appliquent une méthode théorique introduite par \cite{Ledoit11}, qu'ils appellent ``Rotationnaly invariant estimator'', pour debiaiser de manière continue les valeurs propres empiriques et ils montrent que la méthode semble plus robuste que celles du ``Clipping'’ ou du ``Shrinkage'' qui sont bien documentées par \cite{Ledoit01,Ledoit03}). \cite{Allez12} modélisent l'impact du bruit sur le premier vecteur propre et montre que ce dernier tourne légèrement autour d'un vecteur fixe. L'angle de rotation dépend du ratio entre la première valeur propre et les autres. En appliquant ce modèle aux autres vecteurs propres, on comprend qu'ils tournent aussi autour d'axes fixes mais avec un angle de rotation bien plus important. Ils sont ainsi très bruités  ce qui explique la difficulté à les interpréter.

L'ACP avec une contrainte linéaire est une alternative aux filtres issus de la théorie de la matrice aléatoire pour éliminer le bruit de mesure et est entièrement résolu depuis longtemps  (\cite{Golub73}). Dans ce cas les vecteurs propres sous contrainte appartiennent tous au sous espace solution de la contrainte : les vecteurs propres sous contrainte sont simplement les vecteurs propres d'une matrice qui a été réduite et débruitée.  Toute la difficulté est de définir les facteurs formant le sous espace contraint pour que les contraintes n'impactent principalement que le bruit des valeurs propres.  Pour cela, il est possible de s'inspirer de la littérature de l'Asset Pricing décrite dans la section \ref{Assetpricing}) et des modèles multifactoriels décrite dans la section \ref{mutifact}). 

\section{Asset pricing}
\label{Assetpricing}
\cite{Fama65} a abouti  à la théorie des marchés efficients, selon laquelle, les prix suivent des marches aléatoires. Puis \cite{Sharpe64} dérive le MEDAF à partir d'hypothèses plus ou moins réalistes, comme l'absence de coût de transaction et la rationalité des investisseurs. Selon le MEDAF, l’espérance des rendements doit être théoriquement proportionnel au beta, seul risque qui n'est pas diversifiable et qui doit être rémunéré. Depuis 1970, différentes anomalies ont été observées par rapport à cette théorie. Les facteurs classiques de \cite{Fama92,Fama93} sont investis à l'achat sur le top 20 \%, selon le critère financier étudié, et investis à la vente sur le bottom  20\%. Ces facteurs peuvent capturer une anomalie par rapport à la théorie des marchés efficients s'ils génèrent des gains significativement différents de zéros. La construction top 20 \%  bottom  20\% est clairement sous optimale, selon \cite{Asness13}, mais reste paradoxalement la référence dans le domaine de l'Asset Pricing. La régression de \cite{Fama73} est la méthode la plus utilisée pour mettre en évidence des anomalies par rapport au MEDAF. Plusieurs modèles ont été développés pour fournir une interprétation économique aux nombreuses anomalies et pour améliorer le MEDAF. \cite{Fama93}  ont proposé un modèle à trois facteurs pour modéliser les espérances des rendements. \cite{Harvey15} ont listé 316 facteurs potentiels censés capturer une anomalie à partir de 313 articles depuis 1967. Selon eux, la plupart des facteurs peuvent être le fruit du data mining et ne seraient pas robustes. La plupart de ces facteurs se recoupent c'est pourquoi une vingtaine peut suffire mais le niveau de significativité pour caractériser les anomalies ne fait pas consensus. Les travaux académiques ont d'abord retenu les critères financiers tels que la ``Capitalization'', le ``Price Earning Ratio'', le ``Cash Flow'', le ``Book to Market'', la croissance et le ``Momentum''. Par exemple les actions de petites capitalisations tendent à surperformer (\cite{Banz81}). La volume moyen semble plus adéquate que la taille pour \cite{Ciliberti17}. Une autre anomalie importante est la prime ``Value'': les entreprises ``Value'' tendent à surperformer les entreprises de croissance  (\cite{Fama98}). La profitabilité proche du ``Cash Flow'' est aussi une variable explicative significative de l’espérance des rendements (\cite{Fama15}). L'anomalie  ``Low Volatility'' ou ``Low Beta'' ont aussi été révélées (\cite{Jordan13, Fu09, Ang06}). L'anomalie la plus populaire reste le ``Momentum'': les actions qui ont surperformé auront tendance à continuer à surperformer (\cite{Jegadeesh93}). Les anomalies sont directement exploitées dans la gestion d'actifs dont les strategies sont décrites dans la section \ref{ptfmng}. \cite{Asness13}  expliquent ainsi qu'une stratégie de base d'investissement et très populaire simplement allouée en partie sur le ``Momentum'' et sur l'anomalie ``Value'' permet d'atteindre un Sharpe ``in the sample'' supérieur à 1.  

Les théories financières pour justifier de telles primes de risque alternatives  (manque de liquidité, asymétrie)  sont remises en cause car les anomalies ont tendance à disparaitre une fois publiées. \cite{McLean15} ont plusieurs explications alternatives: le biais ``in the sample'' avec le problème de la suroptimisation ou l'adaptation des marchés.

A ma connaissance aucune étude ne s'est encore intéressée à la mise en évidence des autocorrélations des rendements des facteurs de risque qui pourrait constituer une inefficience plus subtile et plus robuste des marchés financiers. Une explication est que les autocorrélations sont trop difficiles à caractériser de manière significative. Des articles existent mais la significativité et la robustesse de leurs résultats ne sont pas convaincants. Ainsi \cite{Hodges17} cherchent des prédicteurs des facteurs dans diffèrents régimes économiques et différentes conditions de marché. Ils trouvent que l'utilisation d'une combinaison d'indicateurs sur le cycle économique, la valorisation, la tendance et la dispersion serait plus efficace que l'utilisation d'indicateurs individuels.

\section{Processus stochastique}
L'instabilité de la matrice de corrélation de population a d'abord été modélisée par des modèles de diffusion pour évaluer des produits dérivés (\cite{Gauthier11}). Les modèles théoriques ont été ajustés pour retrouver les prix des produits dérivés sans chercher à connaitre la réalité de la dynamique de la matrice de corrélation empirique car cette dernière est difficilement mesurable avec la précision recherchée: les modèles ARCH ont été initialement développés pour décrire l'heteroscedasticité des variations de l'inflation (\cite{Engle82}) mais ont ensuite été utilisés pour modéliser la dynamique de la volatilité des actions pour évaluer des options (\cite{Duan95}). Des modèles de type ``Dynamic Conditional Correlation'' (DCC GARCH, \cite{Engle02,Engle16}) ont étendu le modèle GARCH à une dimension et ont été développés pour modéliser la dynamique des corrélations et des volatilités.   De la même façon le processus introduit par \cite{Cox85}, qui est très populaire en finance pour décrire la dynamique des taux d'intérêt et de la volatilité des actions pour évaluer des produits derivés, a aussi été étendu à partir de la diffusion Feller pour modéliser la dynamique des covariances:  les processus de Wishart généralisent à plusieurs dimensions la diffusion de Feller. \cite{Gourieroux2007} introduit ainsi un terme de retour vers la moyenne au processus de Wishart en le rendant stationnaire et généralise  le processus de \cite{Cox85}. \cite{Fonseca2008} généralisent de la même manière le modèle d'\cite{Heston93} pour valoriser les options multi asset.  Un processus de Wishart peut être vu comme le carré de Browniens ou dans sa version stationnaire d'Ornstein-Uhlenbeck. \cite{Cuchiero11} analysent les fondations des processus stochastiques affines continus sur l'univers des matrices de covariance motivé par l'utilisation de tels modèles pour valoriser des options multi-asset ou pour décrire les intensités de défauts.  
\cite{Bru91} dérive les équations stochastiques pour décrire la dynamique de la matrice et la dynamique des valeurs propres. D'autres matrices aléatoires sont aussi très étudiées, comme les matrices gaussiennes dont la distribution des valeurs propres suit la loi circulaire de Wigner.  \cite{Ahdida13} s'intéressent à des matrices de corrélations aléatoires à travers la diffusion de Wright-Fisher pour modéliser les corrélations des actions.   Des algorithmes ont aussi été implémentés pour générer des marches aléatoires parmi les matrices de rotation. Cela permet de décrire la diffusion des vecteurs propres de la matrice de corrélation. Ainsi la marche aléatoire de \cite{Kac1959} est un algorithme assez efficient mais il ne contient pas de retour vers la moyenne, si bien qu'au bout d'un certain temps la matrice n'a plus aucun lien avec la matrice initiale. 

D'autres phénomènes assez fins, comme l'effet de levier restent mal modélisés par les modèles de la littérature. Ainsi des versions asymétriques des modèles type DCC GARCH ont été développées pour tenir compte de l'effet de levier. Malgré une littérature conséquente sur l'effet de levier (quand les prix baissent, la volatilité augmente, selon \cite{Black76,Christie82,Campbell92,Bekaert00,Bouchaud01}), aucun ne s'intéresse à la réalité et la complexité du phénomène bien décrite dans \cite{Bouchaud01}. De nombreux papiers rapportent que les beta, sensibilité des prix des actions aux variations de l'indice, peuvent varier (\cite{Blume71, Fabozzi78,Jagannathan96,Fama97,Bollerslev98,Lettau01,Lewellen06,Ang07}) sans établir de relation précise entre entre l'effet de levier  et l'augmentation des beta. Les actions à fort effet de levier sont plus exposées à un beta instable  (\cite{Galai76,DeJong85}). Bien tenir compte de la variabilité des beta est important aussi pour bien tester les modèles d'Asset Pricing. Ainsi \cite{Bali17} prétendent qu'une fois que les beta sont bien estimés à partir d'un modèle DCC GARCH, alors l'anomalie ``Low Beta'' disparait et le MEDAF est alors enfin vérifié empiriquement (le rendement espéré serait bien proportionnel au beta quand il est bien mesuré).

\chapter{Contributions principales}
Le travail de recherche s'est focalisé sur six sujets pointus et s'est décliné sous la forme de six projets d'articles. Les contributions principales pour chacun des six sujets sont les suivantes:
\begin{itemize}
	\item ``Emergence of Correlation of Securities at Short Time Scales''  (chapitre \ref{emergence}) : l'article introduit un modèle multifactoriel de retard, qui reproduit assez fidèlement les mesures de l'effet d'échelle sur les valeurs propres. Le modèle s'inspire du modèle d'impact de \cite{Kyle85}. Le modèle suppose que les transactions sur les facteurs de risque,  impactent le prix des actions avec un certain retard. Je dérive, sous certaines hypothèses, une formule simple pour décrire la dépendance des valeurs propres avec l'échelle de temps. La formule contient deux paramètres pour chaque valeur propre: la valeur propre asymptotique et un temps de relaxation de l'ordre d'1 minute qui traduit  un retard moyen de l'ordre de quelques minutes entre les actions et les facteurs de risque. Ainsi les corrélations apparaissent à partir d’une minute. Toutefois ce retard de quelques minutes continue d'impacter les valeurs propres de la matrice de corrélation des rendements 20 minutes et au-delà à cause d'une loi en puissance qui s'explique par un mécanisme relativement subtile bien que le phénomène sature. L'article identifie donc une inefficience significative du marché, qui pourrait générer des gains dans le cas théorique où les coûts de transactions sont nuls.
	\item ``The Fundamental Market Neutral Maximum Variance Portfolios''  (chapitre \ref{maxvar}): l'article introduit le ``FCL'' d'un portefeuille (ratio entre la variance du portefeuille et la variance du portefeuille dans le cas où les corrélations entre actions seraient nulles). Le ``FCL''  est un concept proche des valeurs propres et a l'avantage de s'appliquer non seulement aux vecteurs propres mais aussi à n'importe quel facteur de risque. Le ``FCL'' serait une mesure idéale pour caractériser la significativité d'un facteur de risque. J'introduis aussi le portefeuille ``fundamental Max variance'' qui optimise le ``FCL'' et qui peut être interprété comme un vecteur propre de la matrice de corrélation sous contrainte pour capturer au mieux un style donnée defini par un critère financier. Je montre que les poids optimaux dépendent directement des classements des actions en fonction de ce critère et suivent une même loi universelle qui s'applique à tous les critères financiers. Je montre que cette optimisation permet de répliquer au mieux la matrice de corrélation à partir de quelques facteurs ainsi que sa dynamique en filtrant le bruit. Je fais le lien entre les différents ``FCL'', les valeurs propres sous contraintes et les valeurs propres empiriques. Je montre enfin que les vecteurs propres principaux de la matrice de corrélation s'investissent sur les facteurs qui ont les ``FCL'' les plus élevés. Les ``FCL'' sont volatiles et sont bien modélisés par des processus d'Orstein-Ulhenbeck avec un temps de relaxation de 60 jours. La composition des vecteurs propres est donc très variable ce qui explique pourquoi leur interprétation est difficile à l'exception du premier. Je montre aussi sous certaines hypothèses que le Sharpe des portefeuilles ``maximum variance'' est optimal théoriquement. Les résultats de ce chapitre ont été obtenus en collaboration avec Stanislav Kuperstein.
		\item ``Time Scale Effect on Correlation at Long Time Horizon''   (chapitre \ref{emergence2}): l'article décrit une forme plus subtile mais plus robuste d'inefficience des marchés financiers que les écarts entre les espérances non conditionnelles des rendements et les prédictions du MEDAF. Il s'agit de l'autocorrélation des rendements des facteurs de risque qui s'explique par l'illiquidité des marchés financiers et par le comportement moutonnier des investisseurs qui ont tendance à acheter les produits qui ont marché. Cette autocorrélation qui n'est pas décrite dans la littérature va rendre les vecteurs propres et les valeurs propres de la matrice de corrélation sensibles à l'échelle de temps. 
	\item ``The Reactive Beta Model''  (chapitre \ref{beta}): l'article  décrit le modèle de levier systématique (la corrélation augmentent lorsque l'indice baisse), spécifique (le beta d'une action augmente lorsque elle sous performe) et d’élasticité (lorsque la volatilité relative augmente le beta augmente). Il ressort qu'une grande partie de la variabilité des beta s'explique par ces phénomènes. L'approche qui consiste à normaliser les rendements pour corriger ces petits phénomènes permet de réduire le bais de certains facteurs (``Momentum'' et ``Low Beta'') par rapport à la régréssion linéaire directe sur les rendements.   Des tests empiriques montrent la supériorité du modèle par rapport à la simple régréssion linéaire. Des simulations Monte-Carlo montrent aussi l'avantage d'un tel modèle par rapport aux méthodes robustes telles que les régressions par quintiles et les modèles de type DCC GARCH symétriques ou asymétriques. Je montre que mon modèle semble le plus adapté à la réalité des marchés car il a été conçu pour s'adapter à des phénomènes bien caractérisés et mesurés. 
	\item ``The Model of Diffusion of the Correlation between Securities'' (chapitre \ref{diffusion}): l'article identifie quelques faits stylisés qui caractérisent la diffusion des vecteurs propres empiriques des marchés. Les vecteurs propres de la matrice à l'instant t voient leur corrélation en utilisant la matrice à l'instant t +$\tau$ augmenter très légèrement avec $\tau$. Je m'intéresse à la distribution des valeurs propres des incréments de la matrice de corrélation qui est différente de la loi demi-cercle de Wigner et de la distrution qui ressemble à un chapeau pointu. Les équations stochastiques standard (Wright-Fisher, Feller) qui simulent directement la matrice de corrélation ainsi que d'autres méthodes simples qui simulent des trajectoires aléatoires de la matrice de rotation autour de la matrice identité avec un terme de retour vers la moyenne pour simuler la diffusion des vecteurs propres ne permettent pas de reproduire la distribution empirique des valeurs propres. La diffusion des FCL, définies dans le chapitre \ref{maxvar}, permet de generer simplement cette distribution. Les résultats de ce chapitre ont été obtenus en collaboration avec Stanislav Kuperstein.
	\item ``Should Employers Pay Better their Employees? An asset Pricing Approach'' (chapitre \ref{remuneration}) : le facteur rémunération est identifié comme un facteur de risque commun significatif grâce au ``FCL`` mesuré qui est significativement supérieur à 1. Le facteur est ainsi aussi significatif que le facteur ``Book'' de Fama et French. Le facteur rémunération révèle aussi une faible anomalie de marché: les entreprises qui payent bien leurs employés ont un risque en commun et tendent à surperformer les autres. L'article remet en cause la méthodologie de Fama et French qui ne serait pas assez fine pour caractériser une telle anomalie. Ainsi il semble très important de maintenir à chaque instant le facteur beta neutre et pas seulement en moyenne seulement pour mesurer l'anomalie.

\end{itemize}

\part{Dissertation doctorale}
\chapter{Emergence of Correlation between Securities at Short Time Scales}
\label{emergence}
\includepdf[pages=-]{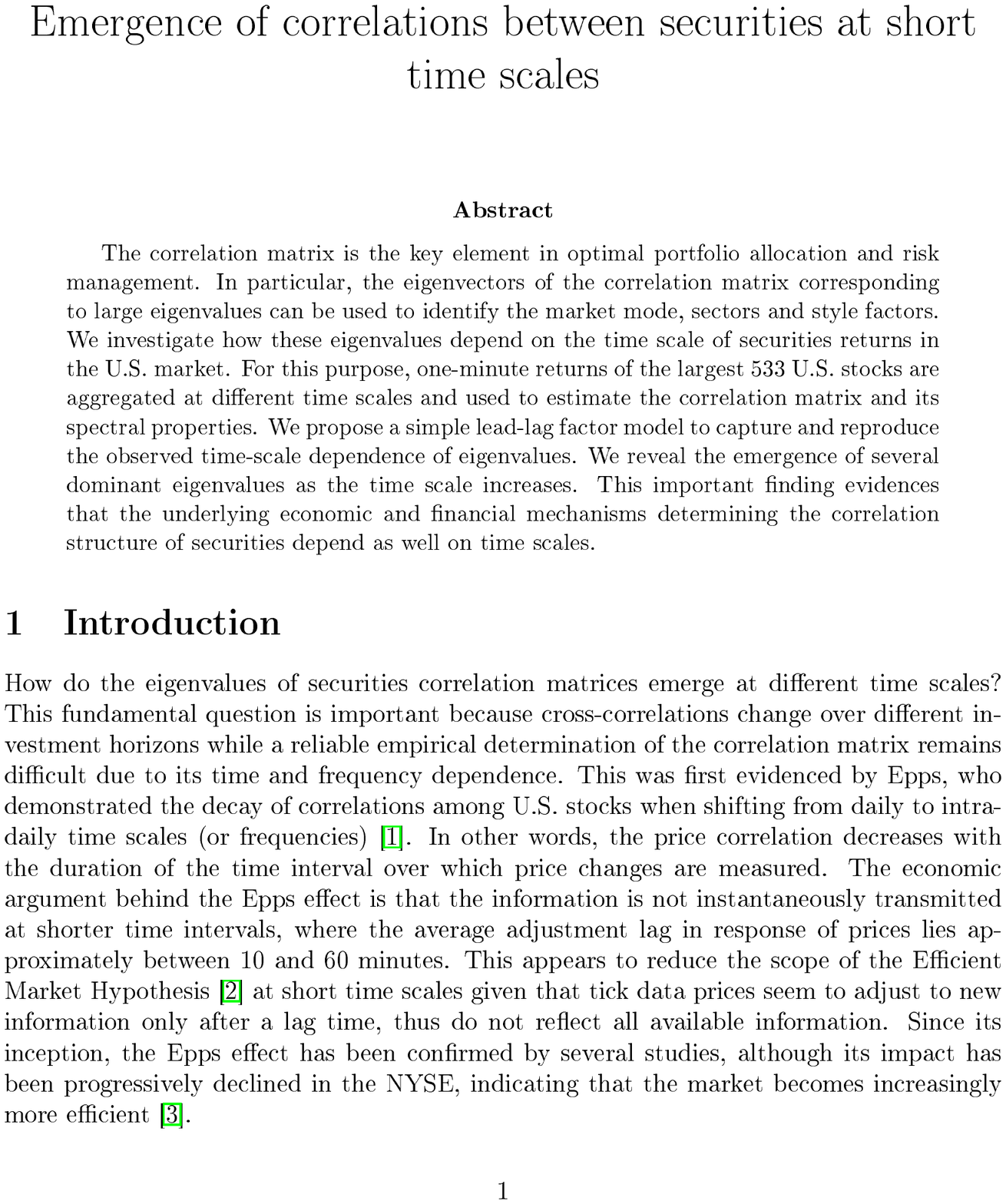}
\chapter{Fundamental Market Neutral Maximum Variance Portfolios}
\label{maxvar}
\footnote{
	The results of this chapter were obtained in collaboration with Stanislav Kuperstein.}
\includepdf[pages=-]{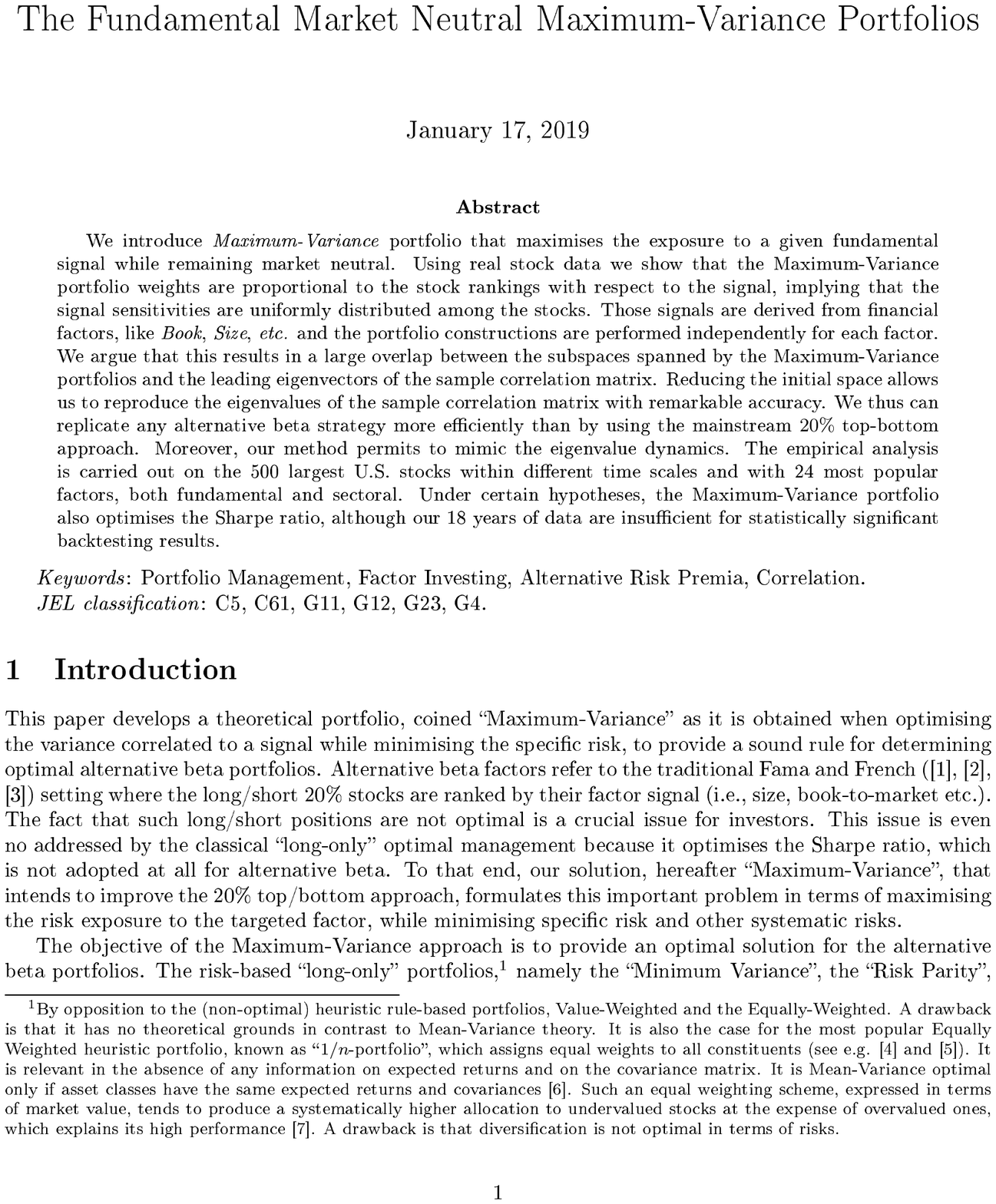}
\chapter{Time Scale Effect on Correlation between Securities at Long Time Horizon}
\label{emergence2}
\includepdf[pages=-]{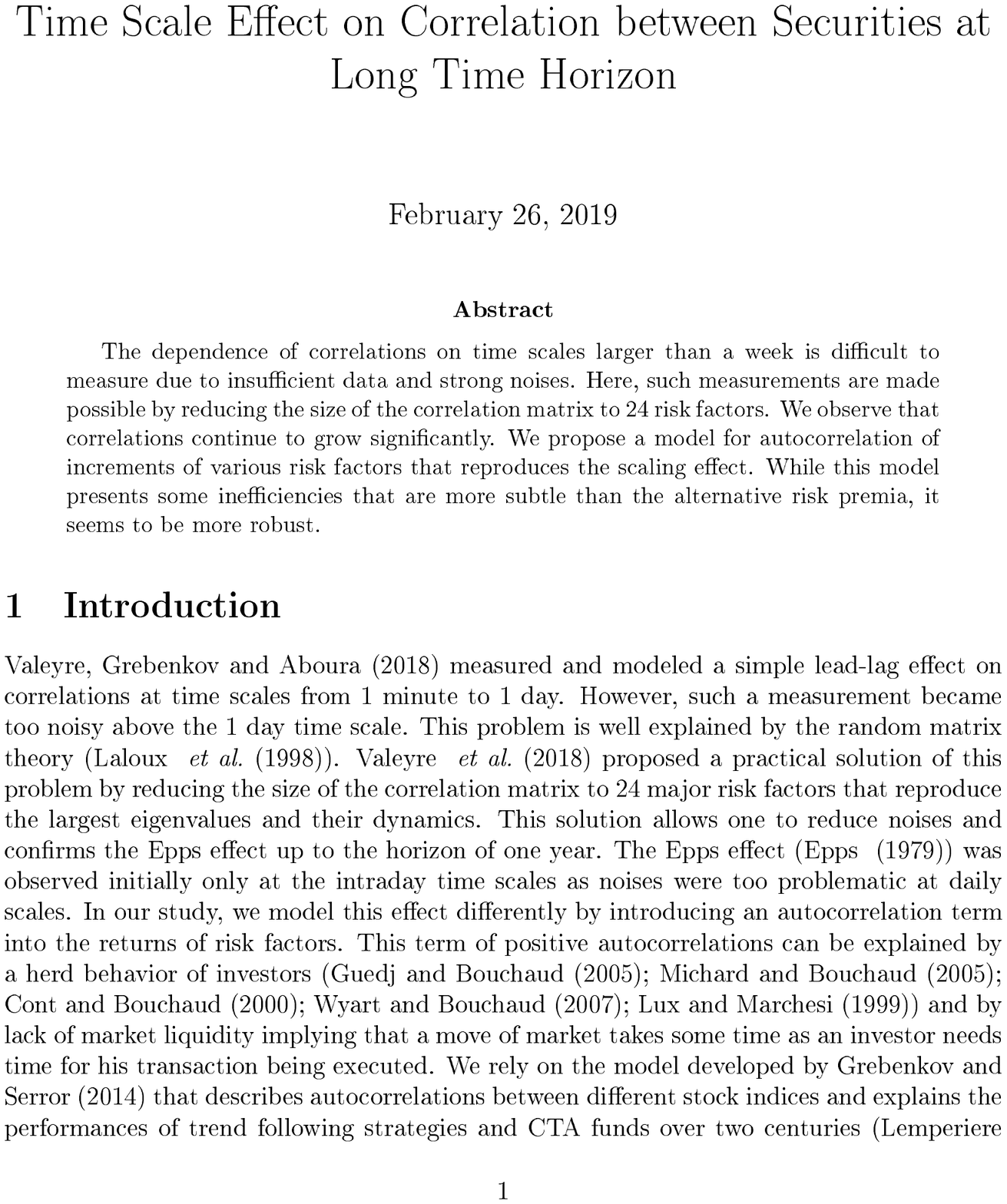}

\chapter{The Reactive Beta Model}
\label{beta}
\includepdf[pages=-]{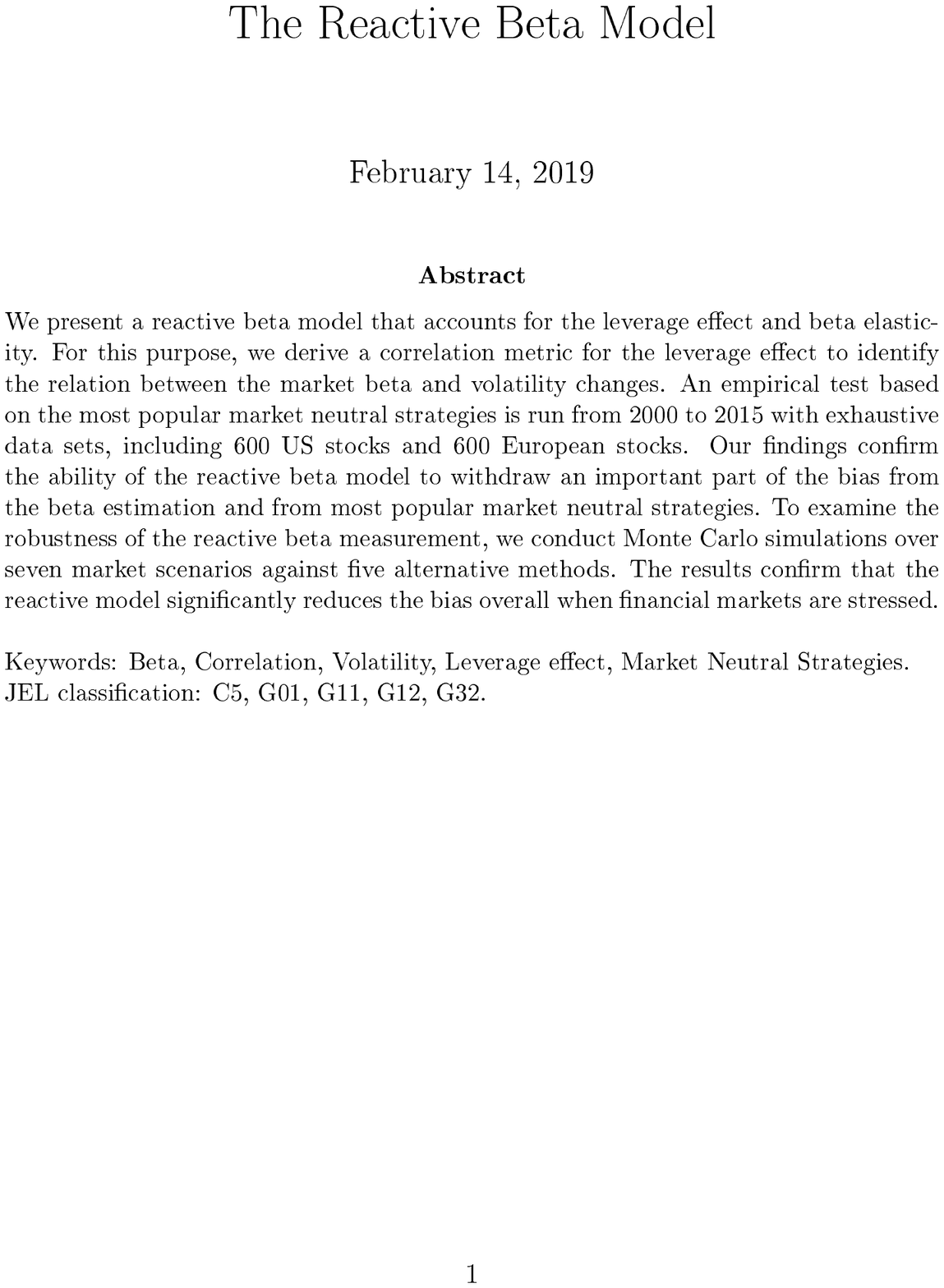}

\chapter{The Model of Diffusion of Correlations between Securities}
\label{diffusion}
\footnote{
	The results of this chapter were obtained in collaboration with Stanislav Kuperstein.}
\includepdf[pages=-]{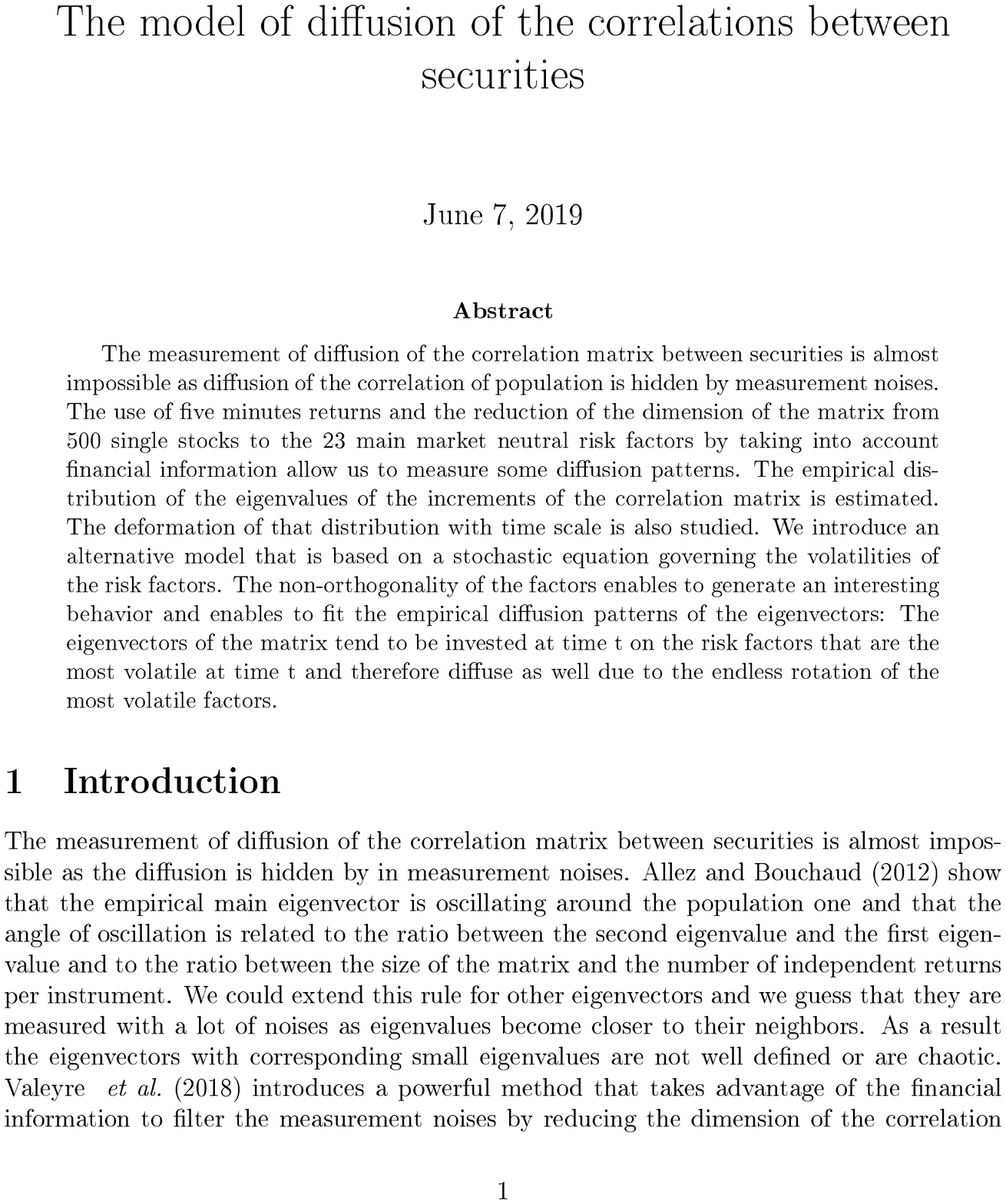}
\chapter{Should Employers Pay Better their Employees? An Asset Pricing Approach}
\label{remuneration}
\includepdf[pages=-]{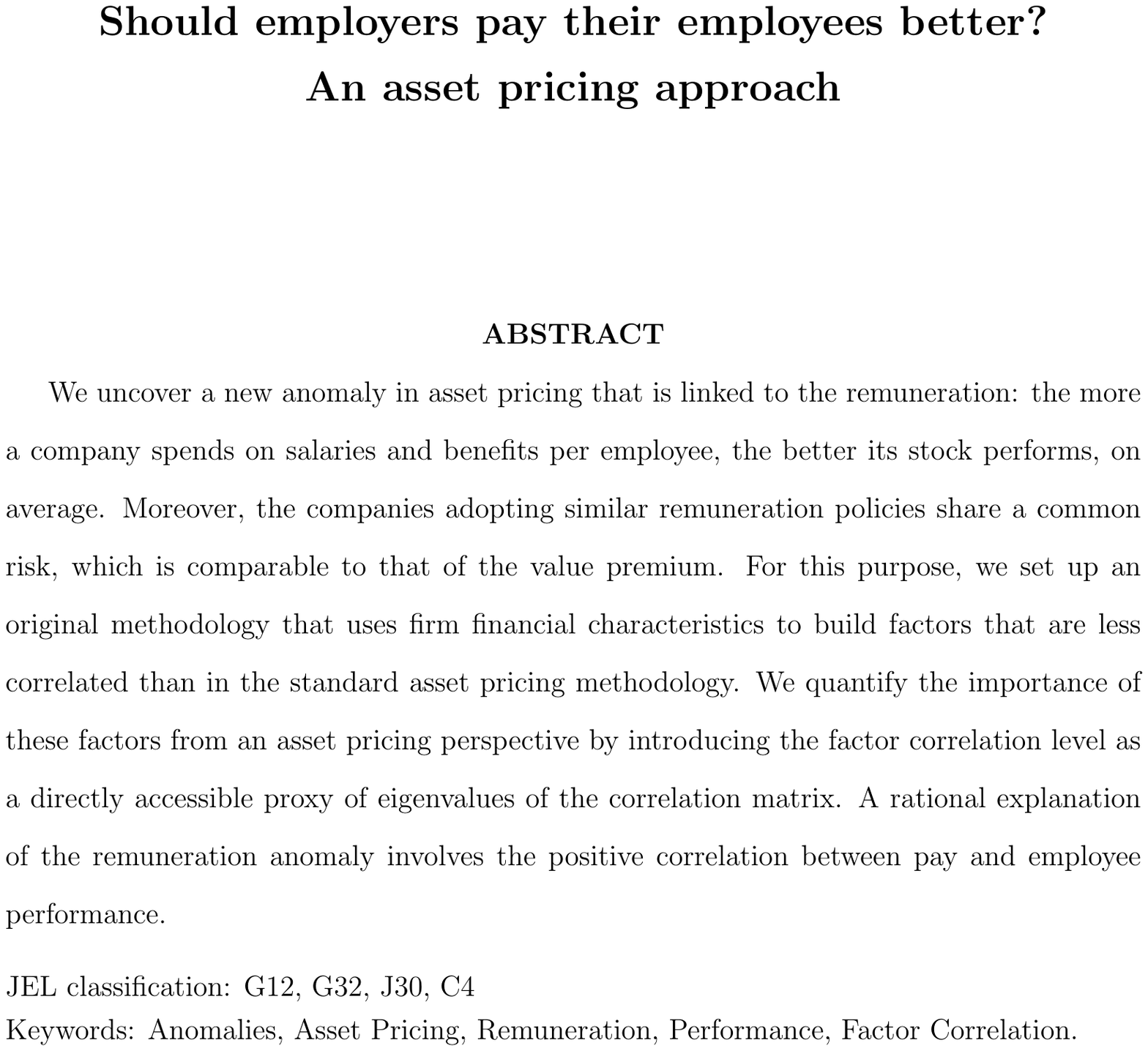}

\part{Conclusion Générale}

Les propriétés empiriques de la matrice de corrélation des rendements des actions ne sont pas bien documentées dans la littérature car elles sont noyées dans le bruit de mesure. L'originalité de la méthode que j'ai introduite  permet de débruiter la matrice de corrélation en profitant de données disponibles en plus des rendements pour contraindre les vecteurs propres. Elle ouvre donc de nouvelles portes. La méthode est particulièrement adaptée aux matrices de corrélation des actions car la première valeur propre est bien plus grande que les autres et de nombreuses données financières sont disponibles (``Book'', ``Capitalization'', ``Cash Flow'', etc.). Le débruitage de la matrice a permis de mettre en évidence de nouvelles propriétés importantes de la matrice de corrélation des actions: \begin{itemize}
	\item l'instabilité des valeurs propres et des vecteurs propres. Ces derniers sont investis en priorité sur les facteurs de risque  les plus importants. L'importance d'un facteur est mesurée à travers le ``FCL'', notion que j'ai introduite. Le ``FCL'' est la variance normalisée d'un facteur de risque et correspond aussi à la moyenne pondérée des valeurs propres par les projections au carré du facteur sur les différents vecteurs propres ;
	\item la diffusion du logarithme des ``FCL'' modélisée par de simples processus d'Orstein-Uhlenbeck semble suffire pour expliquer une grande partie de la diffusion de la matrice de corrélation. Cela permet de retrouver une distribution des valeurs propres des incréments de la matrice de corrélation; 
	\item les poids des facteurs de risque qui optimisent les ``FCL'' sont repartis de manière uniforme ce qui n'est pas compatible avec une distribution aléatoires des vecteurs propres. En effet on aurait à priori attendu une distribution gaussienne des poids qui aurait été naturelle si les vecteurs propres étaient complètement aléatoires. Cela a beaucoup d'applications notamment dans la construction des portefeuilles ``risk premia'' qui sont devenus importants dans l'industrie de la gestion d'actifs. Ces portefeuilles, qui capturent un style donné  sont construits, selon la méthode de Fama et French, avec une fonction ``double Heavyside’’  c'est-à-dire investis à l'achat sur les top 20\% et à la vente sur le bottom 20\% par rapport à un critère donné  (``Book'', ``Capitalization'', ``Momentum'',etc.). Ces portefeuilles peuvent être optimisés avec une règle linéaire compatible avec la distribution uniforme au lieu de la ``double Heavyside’’ de Fama et French. J’ai nommé ces  portefeuilles optimaux  ``Fundamental Market Neutral Maximum Variance Portfolios'' car ces portefeuilles capturent de manière optimale un style donné en minimisant le risque spécifique. Ils ont théoriquement un Sharpe et un ``FCL'' optimaux ;
	\item l'effet d'échelle sur les corrélations avec deux régimes: \begin{itemize}
		
		\item aux petites échelles de temps entre quelques secondes et quelques minutes, un effet de retard de l'ensemble des actions avec un temps de relaxation de quelques minutes explique les petites autocorrélations et l'augmentation des valeurs propres avec l'échelle de temps. J'ai développé un modèle de retard et j'ai dérivé une formule simple qui décrit cette augmentation qui intègre curieusement une loi en puissance. Le modèle reproduit précisément les mesures. Aussi, on peut interpréter les corrélations entre actions comme la conséquence des interactions entre les actions par l'intermédiaire des traders ;
		\item aux grandes échelles de temps entre 1 jour et plusieurs mois une faible autocorrélation est initiée par un manque de liquidité et un comportement moutonnier des acteurs. De la même façon un modèle d'autocorrélation qui inclut des tendances qui suivent un processus d'Ornstein-Ulhenbeck permet de reproduire les augmentations des valeurs propres sur des échelles de temps longues.
	\end{itemize}
	\item l'effet de levier qui est caractérisé par l'augmentation des corrélations et de la  première valeur propre avec la baisse du marché, ne se généralise pas aux autres facteurs de risque. Lorsqu'un facteur chute, son ``FCL'' et les valeurs propres n'augmentent pas. Cela est théoriquement intéressant économiquement dans la mesure où les facteurs de risque alternatifs ne peuvent pas avoir de risque asymétrique sur un horizon de temps long à cause de la loi des grands nombres, s'il n'y a pas d'effet de levier et ne peuvent pas justifier une prime de risque positive. En effet c'est l'effet de levier principalement avec ou sans les queues épaisses des distributions des rendements qui rend la convergence vers la distribution gaussienne très lente en maintenant l'asymétrie. Sans effet de levier les rendements des primes de risque doivent converger plus rapidement vers la distribution gaussienne.  
\end{itemize}

Par ailleurs j'ai aussi étudié finement la dynamique des beta qui est la sensibilité d'une action par rapport aux variations de l'indice, qui est directement liée à  la composition du premier vecteur propre de la matrice de corrélations et qui constitue le paramètre clef de risque. J'ai proposé un modèle réactif avec 3 composants intégrant l'effet de levier spécifique (lorsqu'une action sous performe, son beta augmente), l'effet de levier systématique (lorsque l'indice baisse les corrélations augmentent), l'élasticité des beta (quand la volatilité relative augmente, les beta augmentent). Les trois composants ont été calibrés et testés. J'ai testé le biais du modèle à partir de 4 stratégies ``market neutre’’ de base et j’ai montré la supérioté du modèle par rapport à une simple régression linéaire. J'ai aussi procédé à un test Monte-Carlo qui confirme la supériorité du modèle par rapport aux  méthodes alternatives (``Minimum Absolute Deviation’’, ``Trimean Quantile Regression’’  et ``Dynamic Conditional Correlation’’ avec ou sans asymetrie). 

Enfin j'ai présenté une application très pratique qui présente des implications concrètes pour la gestion d'entreprise en montrant empiriquement que les entreprises qui rémunèrent bien leurs employés partagent une partie significative de leur risque et ont tendance à surperformer. La finesse de methode de mesure permet d'identifier cette anomalie de marché et met en lumière les limitations de la méthode classique de Fama et French. Cette anomalie qui reste néanmoins  relativement faiblement significative semble intuitive et évidente aux professionnels.

\bibliographystyle{te}
\nocite{*}
\bibliography{mainV4}

\newpage 
\begin{changemargin}{-1cm}{-1cm}
{ \setstretch{0.5} 
\begin{footnotesize}
	\textbf{Titre}: Modélisation fine de la matrice de covariance/corrélation des actions
	
	\textbf{Résumé}: Une nouvelle méthode a été mise en place pour débruiter la matrice de corrélation  des rendements des actions en se basant sur une analyse par composante principale sous contrainte en exploitant les données financières. Des portefeuilles, nommés ``Fundamental Maximum variance portfolios'', sont construits pour capturer de manière optimale un style de risque défini par un critère financier (``Book'', ``Capitalization'',etc.). Les vecteurs propres sous contraintes de la matrice de corrélation, qui sont des combinaisons linéaires de ces portefeuilles, sont alors étudiés. Grâce à cette méthode, plusieurs faits stylisés de la matrice ont été mis en évidence dont: i) l'augmentation des premières valeurs propres  avec l'échelle de temps de 1 minute  à plusieurs mois semble suivre la même loi pour toutes les valeurs propres significatives avec deux régimes; ii) une loi ``universelle'' semble gouverner la composition de tous les portefeuilles ``Maximum variance''. Ainsi selon cette loi, les poids optimaux seraient directement proportionnels au classement selon le critère financier étudié; iii) la volatilité de la volatilité des portefeuilles ``Maximum Variance'', qui ne sont pas orthogonaux, suffirait à expliquer une grande partie de la diffusion de la matrice de corrélation; iv) l'effet de levier (augmentation de la première valeur propre avec la baisse du marché) n'existe que pour le premier mode et ne se généralise pas aux autres facteurs de risque. L'effet de levier sur les beta, sensibilité des actions avec le ``market mode’’, rend les poids du premier vecteur propre variables.

	\textbf{Mots clefs}: corrélation, filtre, diagonalisation sous contrainte, modèle multifactoriel, portefeuilles optimaux, gestion d'actifs, diffusion  
	
	\textbf{Discipline}: Sciences Economique/ Gestion de portefeuille
	
		\vspace*{2cm}
	
	\textbf{Title}: Refined model of the covariance/correlation matrix between securities
	
	\textbf{Summary}: A new methodology has been introduced to clean the correlation matrix of single stocks returns based on a constrained principal component analysis using financial data. Portfolios were introduced, namely ``Fundamental Maximum Variance Portfolios'', to capture in an optimal way the risks defined by financial criteria (``Book'', ``Capitalization'', etc.). The constrained eigenvectors of the correlation matrix, which are the linear combination of these portfolios, are then analyzed. Thanks to this methodology, several stylized patterns of the matrix were identified: i) the increase of the first eigenvalue with a time scale from 1 minute to several months seems to follow the same law for all the significant eigenvalues with 2 regimes; ii) a universal law seems to govern the weights of all the ``Maximum variance'' portfolios, so according to that law, the optimal weights should be proportional to the ranking based on the financial studied criteria; iii) the volatility of the volatility of the ``Maximum Variance'' portfolios, which are not orthogonal, could be enough to explain a large part of the diffusion of the correlation matrix; iv) the leverage effect (increase of the first eigenvalue with the decline of the stock market) occurs only for the first mode and cannot be generalized for other factors of risk. The leverage effect on the beta, which is the sensitivity of stocks with the market mode, makes variable the weights of the first eigenvector.

	\textbf{Key words}: correlation, filter, constrained diagonalization, multi factorial model, optimal portfolios, portfolio management, diffusion
	
	\textbf{Discipline}: Economics/ Portfolio Management
\end{footnotesize}

	\vspace*{2cm}

\begin{footnotesize}
\begin{center}
Centre d’Économie de l’Université Paris Nord

U.F.R Sciences Economiques et Gestion

École Doctorale ERASME

Université Paris 13 – Campus Villetaneuse

99 avenue Jean-baptiste Clément

93430 Villetaneuse
\end{center}
\end{footnotesize}
}\par
\end{changemargin}

\end{document}